# Layered Antiferromagnetism Induces Large Negative Magnetoresistance in the van der Waals Semiconductor CrSBr

*Evan J. Telford[Δ], Avalon H. Dismukes[Δ], Kihong Lee, Minghao Cheng, Andrew Wieteska, Amymarie K. Bartholomew, Yu-Sheng Chen, Xiaodong Xu, Abhay N. Pasupathy, Xiaoyang Zhu, Cory R. Dean\*, Xavier Roy\**

E. J. Telford, M. Cheng, A. Wieteska, Prof. A. N. Pasupathy, Prof. C. R. Dean
Department of Physics, Columbia University, New York, NY 10027, USA
E-mail: cd2478@columbia.edu

A. H. Dismukes, Dr. K. Lee, Dr. A. M. Bartholomew, Prof. X. Zhu, Prof. X. Roy
Department of Chemistry, Columbia University, New York, NY 10027, USA
E-mail: xr2114@columbia.edu

Prof. Xiaodong Xu
Department of Physics, University of Washington, Seattle, WA 98195, USA

Dr. Y.-S. Chen
NSF's ChemMatCARS, University of Chicago, Chicago, Il 60439, USA

Δ These authors contributed equally

**Abstract:**

**The recent discovery of magnetism within the family of exfoliatable van der Waals (vdW) compounds has attracted considerable interest in these materials for both fundamental research and technological applications. However current vdW magnets are limited by their extreme sensitivity to air, low ordering temperatures, and poor charge transport properties. Here we report the magnetic and electronic properties of CrSBr, an air-stable vdW antiferromagnetic semiconductor that readily cleaves perpendicular to the stacking axis. Below its Néel temperature, $T_N$ = 132 ± 1 K, CrSBr adopts an A-type antiferromagnetic structure with each individual layer ferromagnetically ordered internally and the layers coupled antiferromagnetically along the stacking direction. Scanning tunneling spectroscopy and photoluminescence (PL) reveal that the electronic gap is $\Delta_E$ = 1.5 ± 0.2 eV with a corresponding PL peak centered at 1.25 ± 0.07 eV. Using**

**magnetotransport measurements, we demonstrate strong coupling between magnetic order and transport properties in CrSBr, leading to a large negative magnetoresistance response that is unique amongst vdW materials. These findings establish CrSBr as a promising material platform for increasing the applicability of vdW magnets to the field of spin-based electronics.**

Main Text:

Materials that combine bulk magnetic order and semiconducting transport properties have received widespread attention for their ability to control both charge and spin carriers, allowing for complete spin polarization of their conduction electrons.[1] By exploiting the spins of electrons as information carriers, instead of their charge, these materials promise to improve the speed, density, and energy efficiency of electronic devices through single-spin transport.[2–4] This makes magnetic semiconductors particularly attractive for device applications that utilize the electronic tunability, spin-polarized transport, and exotic magneto-optical properties characteristic of magnetic semiconductors such as magnetic tunnel junctions and spin field effect transistors.[4–10] For these reasons, semiconductors with strongly coupled electrical properties and magnetic order offer advantages over magnetic metals and insulators but they are comparatively notably scarce.[11] Thus, synthesizing new materials that intrinsically exhibit both semiconducting behavior and magnetic ordering is imperative from a material design perspective to advancing spin-based technologies.

The recent discovery of two-dimensional (2D) magnets from bulk van der Waals (vdW) materials[12–14] provides an ideal platform to understand and ultimately control 2D magnetism, fueling opportunities for atomically-thin spintronic and magneto-optic devices, valleytronics and on-chip communication.[15–19] In addition, 2D magnets can be used to engineer interfacial phenomena in vdW heterostructures through the proximity effect,[18,20–22] including spin superexchange, anomalous Hall effects, and topological superconductivity. The realization of a 2D magnetic semiconductor would be particularly exciting owing to the possibility of taking the increased electronic tunability and unique magnetotransport and magneto-optical properties of bulk magnetic semiconductors down to the few-layer limit. To date, however, the design of vdW materials, from which 2D magnets can be exfoliated, remains a major synthetic challenge underscored by the inadequacies of currently available materials, including low magnetic transition temperatures (45 K and 25 K for few layer CrI$_3$[12] and Cr$_2$Ge$_2$Te$_6$,[14] respectively), extreme air instability, and poor transport properties. In this context, the vdW material CrSBr is expected to be an important milestone: bulk CrSBr has a high antiferromagnetic ordering temperature ($T_N$ ~ 132 K) with a number of theoretical studies predicting the monolayer to have an even higher ferromagnetic ordering temperature ($T_C$ ~ 150 K), semiconducting transport properties, stability under ambient conditions, and gate-tunable magnetic ordering.[1,23,24] Despite its promise, few experimental probes have been reported.[25]

In this work, we synthesized CrSBr and reveal strong coupling between its magnetic ordering and transport properties. We measured magnetotransport down to 5 K and observed large intrinsic negative magnetoresistance (nMR) of up to ~40%, almost 10 times greater than the magnetoresistance in typical metallic magnetic materials[26–29] (< 5%) and more than twice that of the magnetoresistance values in many dilute magnetic semiconductors (~15%).[30,31] Magnetic measurements on single crystals confirm the previously reported magnetic structure: the emergence of an A-type antiferromagnetic (AF) phase below $T_N$ = 132 ± 1 K consisting of ferromagnetic (FM) layers, with an in-plane magnetic easy axis along the crystallographic *b*-axis, coupled antiferromagnetically along the *c*-axis (**Figure 1A,B**).[25] The magnetic structure of CrSBr, coupled with its easily cleavable layered vdW structure, high magnetic ordering temperature, and strong coupling between magnetism and transport make this compound attractive to advance the fields of magnetic semiconductors, 2D magnetism, and nanospintronics.

We grew CrSBr single crystals as millimeter scale shiny black flat needles using a modified chemical vapor transport approach adapted from the original method reported by Beck.[32] Single crystal x-ray diffraction (SCXRD) crystallography reveals the layered vdW structure of CrSBr (**Figure 1A,B** and **Table S1**). Each layer is made of two buckled planes of CrS sandwiched between Br sheets; the CrSBr layers stack along the *c*-axis through vdW interactions.[25,32,33] The space group is *Pmmn* ($D_{2h}$), which consists of a rectangular structure as viewed along the *c*-axis (7.96 Å), with a larger interplanar spacing of the *b*-axis (4.76 Å) than the *a*-axis (3.50 Å). The chemical composition is confirmed by energy dispersive x-ray spectroscopy (**Figure S1**). The crystal symmetry remains unchanged between 15 and 300 K (**Table S1**) and the crystal structure, magnetic and optoelectronic properties are stable under ambient conditions (**Figure S2**). The needle shape of the crystals reflects the in-plane lattice anisotropy, allowing the identification of crystal axes from the morphology and experimentally observed anisotropic Raman peaks (**Figure S3**). A consequence of the large interlayer spacing and vdW stacking is the propensity of the crystals to cleave parallel to the *ab*-planes, providing the ability to mechanically exfoliate CrSBr with Scotch tape and expose pristine surfaces (**Figure 1C**).

We performed electrical transport measurements on bulk CrSBr crystals (~500×200×10 $\mu$m$^3$) as a function of temperature and magnetic field. The devices were fabricated by cleaving the CrSBr crystals to obtain pristine surfaces (**Figure 1C**), mounting them into a non-conducting DIP socket and making electrical contact with silver paint. Scanning tunneling microscopy (STM) imaging of the exposed surface shows the rectangular lattice with interatomic distances consistent with SCXRD (**Figure 1D and S4**). The contacts are arranged in a Hall bar geometry, allowing simultaneous measurement of the longitudinal ($R_{xx}$) and Hall ($R_{xy}$) resistances (details can be found in the Supporting Information). The crystals display an average sheet resistance ($R_S$) of 240 ± 60 kΩ at room temperature, where we use the general definition $R_S = \frac{\rho_{xx}}{t} = R_{xx}\frac{W}{L}$ (where $W$, $L$, and $t$ are the channel width, length, and thickness, respectively). The conductance versus temperature (**Figure 2A**) shows a semiconducting response, with conductance decreasing with decreasing temperature. This is consistent with scanning tunneling spectroscopy (STS) measurements (**Figure 1E**) as well as photoluminescence spectroscopy (**Figure 1F**), which identify an electronic gap $\Delta_E = 1.5 \pm 0.2$ eV and a PL peak centered at $1.25 \pm 0.07$ eV, respectively. The Fermi level is located close to the conduction band edge ($E_C - E_F < 100$ meV) as measured by STM (**Figure 1E**), indicating that the system is electron doped. The measured transport properties also support this claim. At high temperature, the conductance versus temperature is well described by a thermally activated model, $G \propto e^{-(E_C - E_F)/k_B T}$, with $E_C - E_F = 93 \pm 14$ meV (averaged over 4 devices) (**Figure 2A** and **S6**), in excellent agreement with STS measurements. The sign and magnitude of the room temperature sheet carrier density ($n_{2D} \sim 5 \times 10^{13}$ cm$^{-2}$) as determined from the Hall effect confirm electron doping (**Figure S5**).

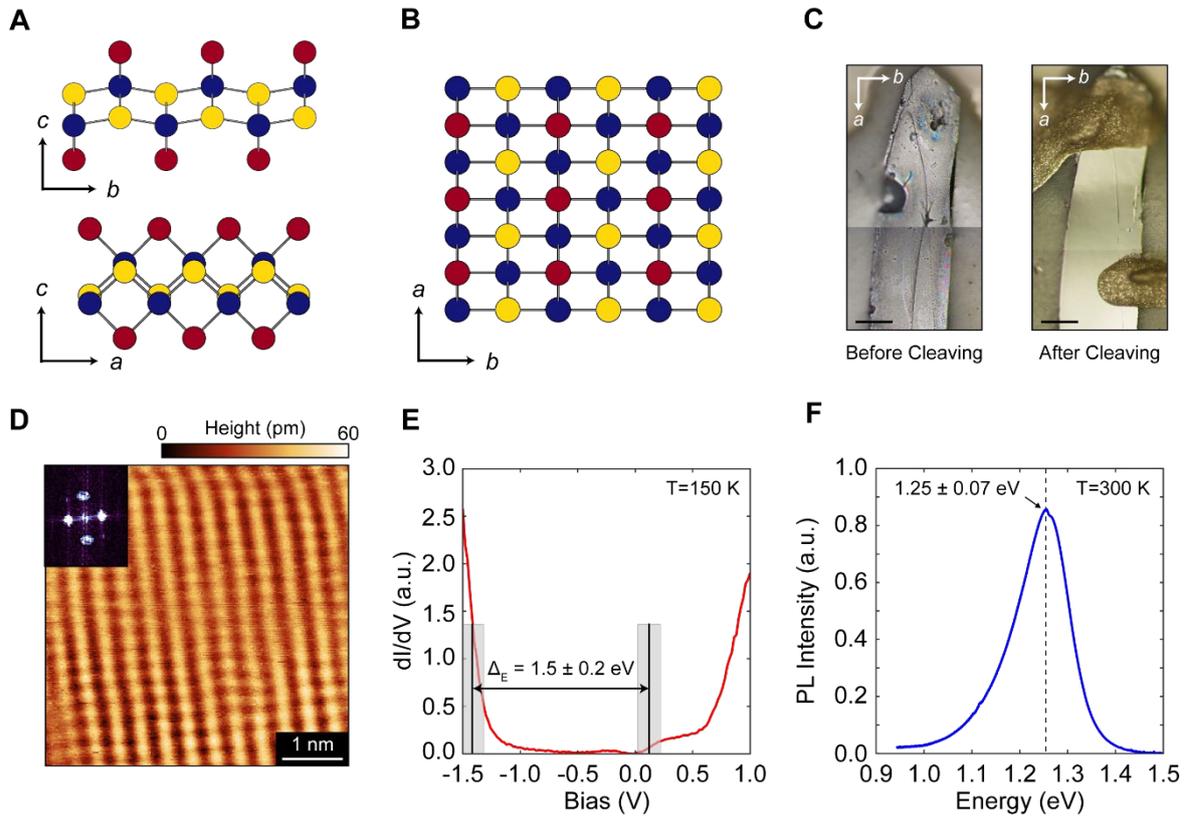

**Figure 1. Crystal structure and semiconducting behavior of CrSBr.** (**A, B**) Crystal structure of CrSBr as viewed along the *a*-axis (**A: top**), *b*-axis (**A: bottom**), and *c*-axis (**B**). Cr, S, and Br atoms are false colored to appear blue, yellow, and red, respectively. (**C**) Optical images of a CrSBr single crystal before (left) and after (right) mechanically cleaving with Scotch tape. The scale bars are 100 μm. Orientation of the crystal axes are given in the inset. (**D**) Scanning tunneling microscopy image of bulk CrSBr along the *c*- axis at $T = 150$ K. STM topographical images were obtained in constant current mode ($V_{bias} = 2$V, $I_{tunneling} = 2$ pA). An FFT of the topography is given in the inset. (**E**) $dI/dV$ versus bias, as measured by scanning tunneling spectroscopy. The gap is extracted by fitting a linear segment on both the left and right side of the STS, then taking the mid points of the linear segments (represented by solid black lines) as the gap edges. The extracted electronic bandgap is given in the inset. The grey boxes demarcate the error in determining the gap edges. **F)** Photoluminescence intensity versus excitation energy at $T = 300$ K. The position of the PL peak is denoted by a black dashed line and given in the inset.

The plot of the conductance versus temperature (**Figure 2A**) displays a sudden change in the slope at $T = 148 \pm 6$ K. We define the transition temperature as the midpoint through the change in slope with the error defined as the width of the transition. The position of the kink in the conductance and the increase in the slope of conductance versus temperature below the kink are observed in all measured samples (**Figure S6**). **Figure 2B** shows the corresponding temperature dependence of the magnetic susceptibility ($\chi$) of CrSBr along each crystallographic axis, as measured by SQUID magnetometry. The sharp cusp at $T = 132 \pm 1$ K identifies the Néel temperature ($T_N$), signaling that the system transitions to an AF phase[34] below this temperature. $T_N$ is closely matched to the transition in conductance versus temperature, indicating that this kink originates from the onset of the AF phase.[28,35,36] Above $T_N$, $\chi$ follows the Curie-Weiss law ($\chi = \chi_0 + \frac{C}{T - \theta_{CW}}$) (where $\chi_0$ is a temperature independent term including core diamagnetism and the background signal from the sample mount, $C$ is the Curie constant, and $\theta_{CW}$ is the Weiss constant), demonstrating bulk paramagnetic (PM) behavior (**Figure S7**).[34] The extracted Weiss constants are larger than $T_N$ and positive ($\theta_{CW}$ = 185, 164 and 184 K for the *a*-, *b*-, and *c*-axes, respectively), signifying strong local FM interactions in the system (**Table S2**). We note that in all devices, $\chi$ versus $T$ manifests a small kink at $T \sim 30$ K, the origin of which is unclear at present but will be the subject of future investigation. Along the *b*-axis, the magnetic phase transition in the $\chi$ versus $T$ plot is sharpest and the decrease of the low-temperature $\chi$ is more prominent than along the *a*- or *c*-axes. Together, these observations indicate that the crystallographic *b*-axis is the easy magnetic axis. The *a*- and *c*-axes are the magnetic intermediate and hard axes, respectively.[25]

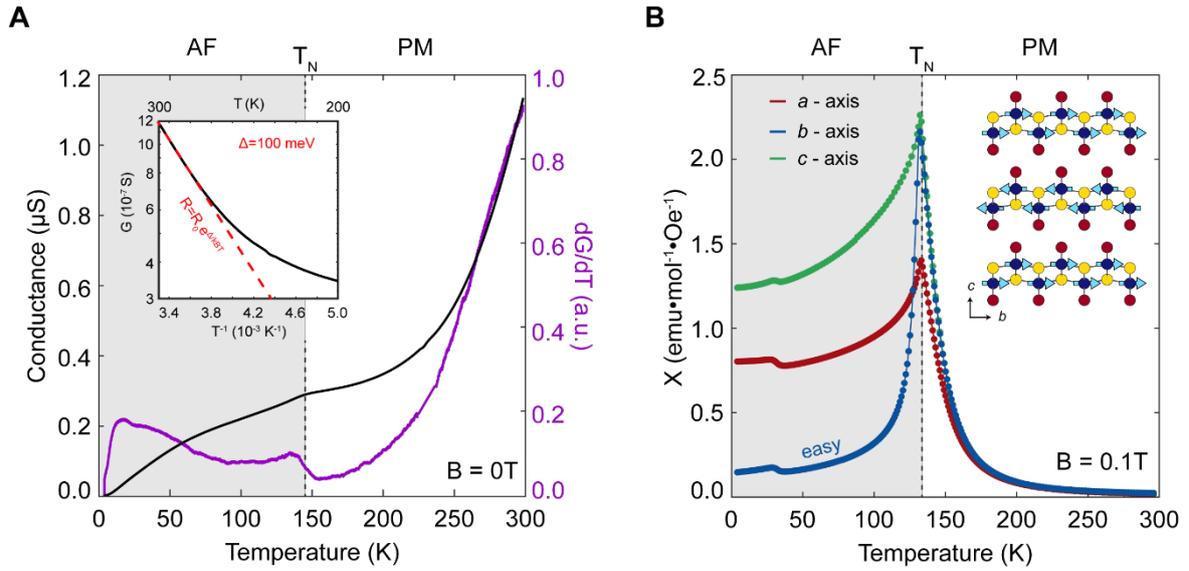

**Figure 2. Electrical transport and magnetic properties of CrSBr.** (**A**) Conductance (black solid line) and the derivative of conductance (purple solid line) versus temperature at zero magnetic field. The inset plots conductance on a log scale versus inverse temperature. An exponential fit to the data is given by a red dashed line. The extracted transport gap is given in the inset. (**B**) Magnetic susceptibility versus temperature along the *a*-axis (solid red dots), *b*-axis (solid blue dots), and the *c*-axis (solid green dots). Inset shows the magnetic ordering of the Cr spins in the AF state. The easy axis (*b*-axis) is denoted. The data was collected with a magnetic field of 100 mT. In (**A**) and (**B**), the AF and PM phases are denoted with grey and white regions, respectively.

**Figure 3A-C** presents the magnetoresistance ratio (defined as $MRR = \frac{R(B)-R(B=0)}{R(B=0)} \times 100$) versus applied magnetic field ($B$) along each crystallographic axis as a function of temperature. Above $T_N$, we observe a small but finite negative magnetoresistance versus $B$. This is characteristic behavior of a PM state, typically attributed to quenching of spin disorder upon applying an external magnetic field.[28,37] At $T < T_N$, the negative magnetoresistance (nMR) in magnetotransport is strongly enhanced along all crystallographic axes. With increasing $B$, the nMR is followed by a saturation of the magnetoresistance when the spins are completely aligned with $B$ in the fully polarized (FP) state.[28,35,36] The field at which the magnetoresistance begins to saturate is denoted as the saturation field ($H_S$). This behavior can be explained by the fact that the AF phase suppresses interlayer tunneling due to adjacent spins having opposite magnetization. In the FP state all spins are aligned, restoring interlayer tunneling, leading to a decrease in the overall sample resistance.[38–40] In addition, the magnetoresistance curves develop hysteresis along every field direction below $T_N$ which increases in magnitude as temperature decreases (**Figure S8**), consistent with the formation of magnetic domains. Below ~ 30 K, the magnetotransport shows strong anisotropy with the emergence of a non-monotonic response along the *a*- and *c*-axes. Additionally, at this temperature the initially negative magnetoresistance changes sign and becomes positive with further increasing $B$, until it saturates at $H_S$ to a constant value when the FP state is reached. Similar low temperature non-monotonic behaviors observed in other systems[36] have been interpreted to arise from rotating off-axis in-plane domains or secondary magnetic ordering.[36,37,41]

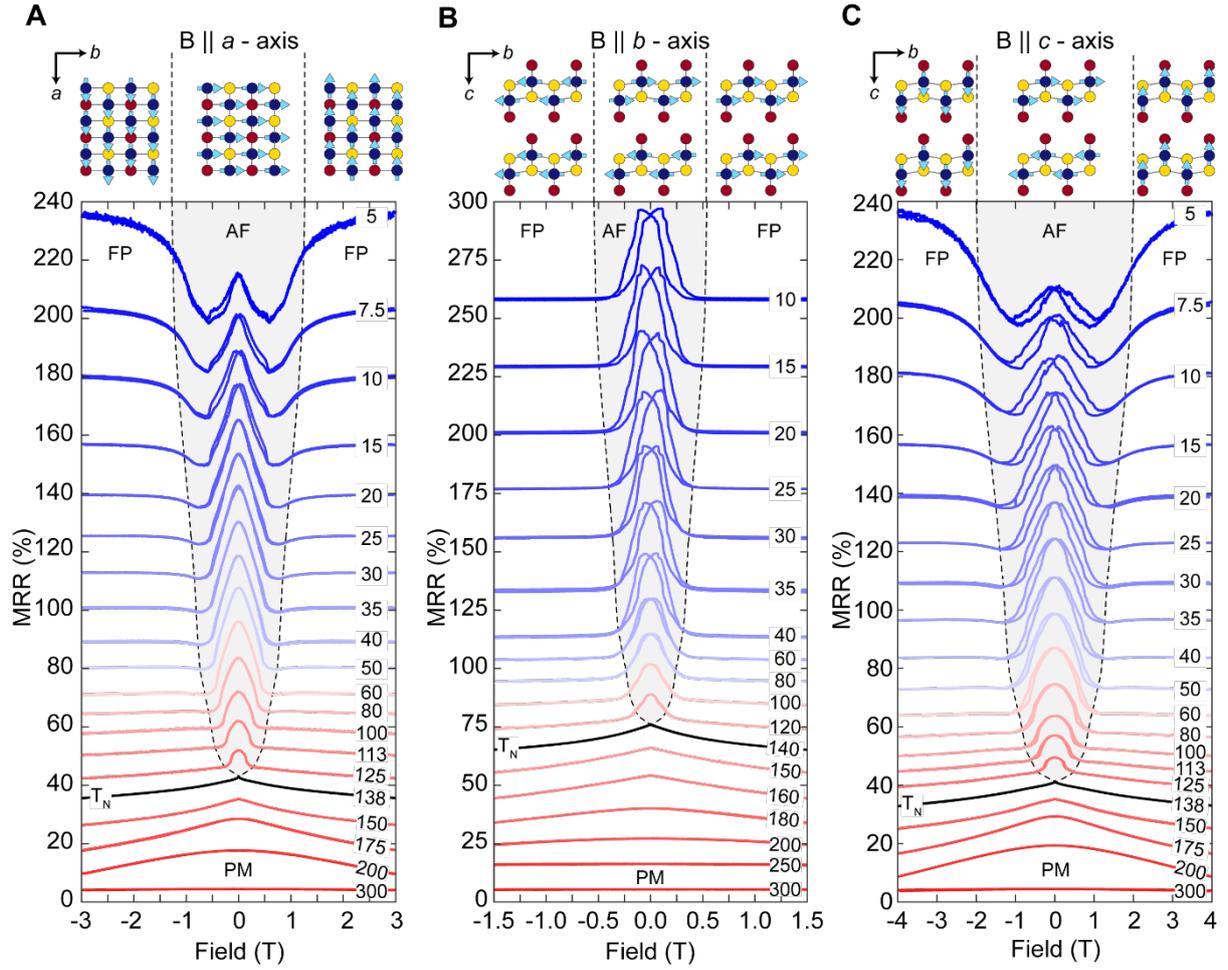

**Figure 3. Magnetotransport properties of CrSBr.** Magnetoresistance ratio (defined as $MRR(B) = \frac{R(B)-R(B=0)}{R(B=0)} \times 100$) versus magnetic field at various temperatures with the field oriented along the *a*-axis (**A**), *b*-axis (**B**), and *c*-axis (**C**). Both forward and backward magnetic field scans at each temperature are presented. The curves are offset for clarity. The solid black lines represent curves taken at temperatures near $T_N$. The AF, FP, and PM phases are labelled, and the phase boundary is denoted by dashed black lines. Schematics showing the orientation of the spins in the AF and FP state are given above each plot.

To further our understanding of the magnetotransport, we measured the magnetization (*M*) of CrSBr as a function of the applied magnetic field along each crystal axis via SQUID magnetometry (**Figure 4A-C**). Along the *b*-axis, *M* sharply transitions from zero to the saturation magnetization, typical of a first order AF-FP spin-flip transition (**Figure 4B**).[25,34] The *M* versus *B* curves along the *a*- and *c*-axes (in-plane intermediate and out-of-plane hard magnetic axes, respectively) exhibit a continuous increase of *M* from zero to the saturation magnetization, indicating that the spins are progressively canting to align with *B* (**Figure 4A,C**). The sharpness of the AF-FP transition along the *b*-axis confirms that it is the magnetic easy axis.[25] The saturation magnetic fields determined from the magnetization curves correlate well with $H_S$ extracted from the magnetoresistance data for each field direction. From the combined magnetotransport and magnetization measurements, we conclude that the electrical properties of CrSBr are strongly coupled to the magnetic order; the observed magnetotransport properties are a direct result of the layered antiferromagnetic structure of CrSBr.

**Figure 4D,E** summarizes the magnetoelectronic properties of CrSBr extracted from magnetotransport measurements. **Figure 4D** presents the temperature dependence of $H_S$ along each field direction. $H_S$ is extracted by plotting the second derivative of *MRR* ($\frac{d^2}{dB^2} MRR(B)$) versus *B* and *T* and extracting contour lines near $\frac{d^2}{dB^2} MRR(B) = 0$ (**Figure S9**). $H_S$ is largest along the magnetically hard *c*-axis and smallest along the easy *b*-axis. At low *T*, it decreases linearly with increasing temperature and then asymptotically approaches zero as we come close to $T_N$. The zero-temperature $H_S$ values ($H_{S0}^a$ = 1.17 T, $H_{S0}^b$ = 0.58 T, $H_{S0}^c$ = 2.00 T) are consistent with magnetization measurements (**Figure 4A-C**) and the corresponding $T_N$ (~ 140 K) is consistent with the zero-field conductivity measurements (**Figure 2A**).

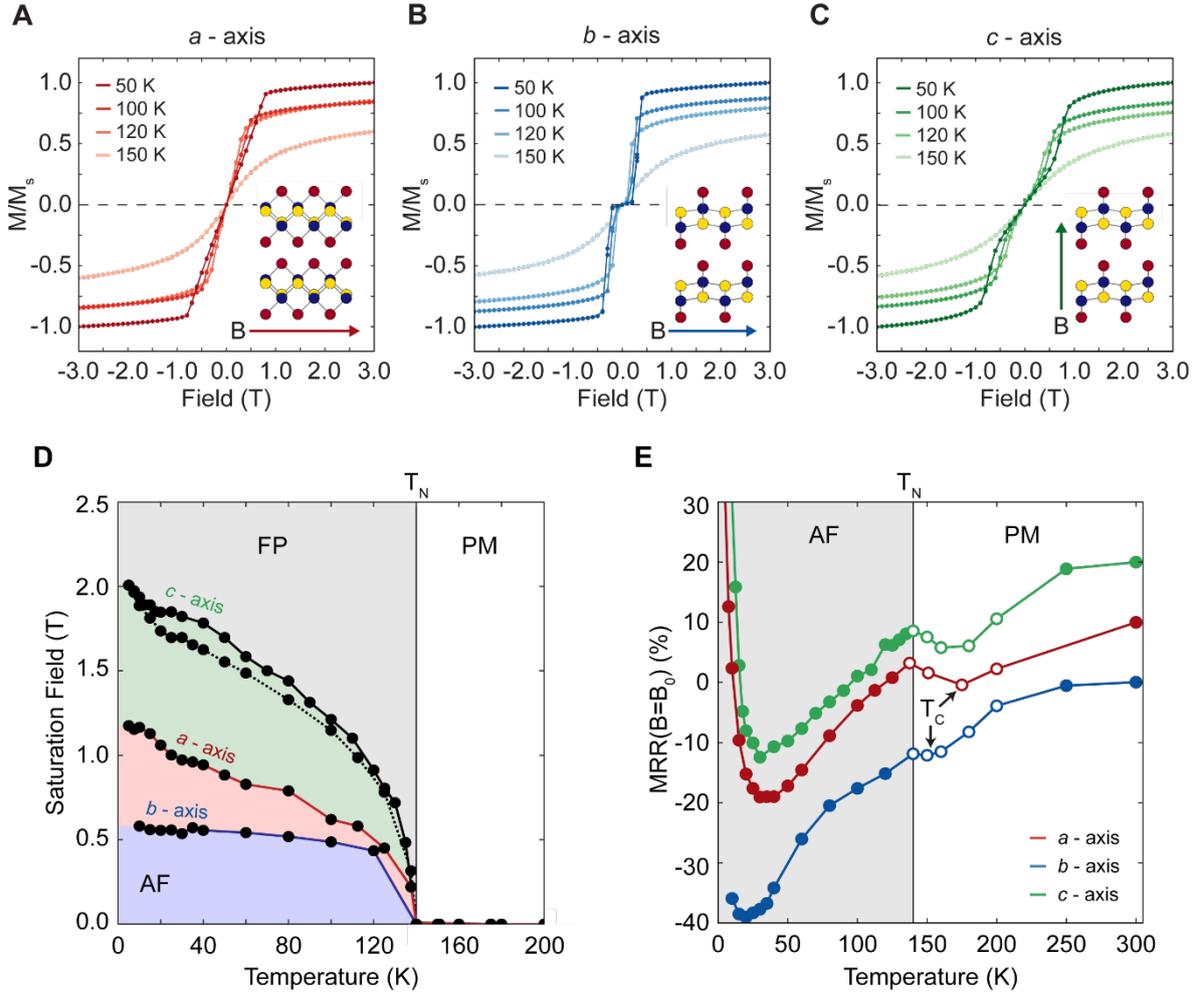

**Figure 4. Saturation fields and magnetoresistance ratios versus temperature.** (**A-C**) Magnetization versus magnetic field at various temperatures along the *a*-axis (**A**), *b*-axis (**B**), and *c*-axis (**C**). *M* is normalized to the saturation magnetization at *T* = 50 K. Insets of **A-C** show the magnetic field direction relative to the crystal structure. (**D**) Saturation magnetic field versus temperature along the *a*-axis (red region), *b*-axis (blue region), and *c*-axis (green region) as measured by extracting the crossover point from the positive/negative magnetoresistance regime to the saturated regime. (**E**) *MRR* at a fixed magnetic field $MRR(B=B_0) = \frac{R(B=B_0)-R(B=0)}{R(B=0)} \times 100$) versus temperature along the *a*-axis (red dots), *b*-axis (blue dots), and *c*-axis (green dots). Each curve is offset from one other by 20% for clarity. All curves have an *MRR* = 0% at room temperature. The magnetic fields at which the fixed-field *MRR* is calculated are $B_0$ = 3, 2, and 4 T for fields along the *a*-, *b*-, and *c*-axes, respectively. A clear kink in the fixed-field *MRR* versus temperature above $T_N$ is demarcated along each axis. The open colored circles denote *MRR* curves versus *B* that were not saturated at the chosen $B_0$ value, indicating the *MRR* may continue to decrease for larger *B*.

In **Figure 4E**, we plot the fixed-field *MRR* defined as *MRR(B=B₀)* = $\frac{R(B=B_0)-R(B=0)}{R(B=0)} \times 100$ (where $B_0$ = 3, 2, and 4 T for the *a*-, *b*-, and *c*-axes, respectively) along each crystallographic axis. Along all axes, the fixed-field *MRR* increases significantly below $T_N$ up to a maximum of –29.3%, –39.2%, and –32.7% at a temperature of ∼ 30 K for the *a*-, *b*-, and *c*-axes, respectively. Such giant intrinsic fixed-field *MRR* is nearly 10 times larger than typical magnetic metals[26–29] and double that of many dilute magnetic semiconductors.[30,31] Above $T_N$, the fixed-field *MRR* is non-zero and negative, manifesting a local minimum between 150 and 170 K. This temperature range is consistent with the Weiss constants extracted from the magnetic susceptibility data ($\theta_{CW}$ = 185, 164 and 184 K for the *a*-, *b*-, and *c*-axes, respectively) (**Table S2**) and signals the emergence of a hidden magnetic phase above $T_N$ characterized by ferromagnetically ordered layers, with no long range magnetic order between the layers.[42]

In summary, we demonstrate CrSBr to be a robust antiferromagnetic semiconductor with exceptionally strong coupling between magnetic and electronic properties manifested through the observation of nMR up to ∼40%. STM and STS studies combined with PL measurements reveal that CrSBr is a direct gap semiconductor with an electrical gap $\Delta_E$ = 1.5 ± 0.2 eV and a PL peak centered at 1.25 ± 0.07 eV. Magnetic measurements on single crystals show the emergence of an AF phase consisting of FM layers coupled antiferromagnetically along the stacking direction below $T_N$ = 132 ± 1 K with a magnetically easy axis parallel to the crystallographic *b*-axis. In magnetotransport, we observe strong coupling between the magnetic order and the electrical properties through the observation of giant nMR below $T_N$ along all crystallographic axes. The magnetotransport properties of CrSBr, in combination with the ability to easily cleave perpendicular to the stacking axis, establish its unique potential for advancing the fields of 2D magnetism and nanospintronics.

**Acknowledgements**
EJT and AHD contributed equally to this work. We thank Johannes Beck for helpful discussion on the synthesis. Research on magnetotransport properties of vdW magnetic semiconductors is supported as part of Programmable Quantum Materials, an Energy Frontier Research Center funded by the U.S. Department of Energy (DOE), Office of Science, Basic Energy Sciences (BES), under award DE-SC0019443. AD is supported by the NSF graduate research fellowship program (DGE 16-44869). Low temperature SCXRD was performed at NSF's ChemMatCARS at Argonne National lab, supported by the Divisions of Chemistry (CHE) and Materials Research (DMR), National Science Foundation, under grant number NSF/CHE-1834750. Use of the Advanced Photon Source, an Office of Science User Facility operated for the U.S. Department of Energy (DOE) Office of Science by Argonne National Laboratory, is supported by the U.S. DOE under Contract No. DE-AC02-06CH11357.

Supporting Information

**Layered Antiferromagnetism Induces Large Negative Magnetoresistance in the van der Waals Semiconductor CrSBr**

*Evan J. Telford$^\Delta$, Avalon H. Dismukes$^\Delta$, Kihong Lee, Minghao Cheng, Andrew Wieteska, Amymarie K. Bartholomew, Yu-Sheng Chen, Xiaodong Xu, Abhay N. Pasupathy, Xiaoyang Zhu, Cory R. Dean\*, Xavier Roy\**

E. J. Telford, M. Cheng, A. Wieteska, Prof. A. N. Pasupathy, Prof. C. R. Dean
Department of Physics, Columbia University, New York, NY 10027, USA
E-mail: cd2478@columbia.edu

A. H. Dismukes, Dr. K. Lee, Dr. A. M. Bartholomew, Prof. X. Zhu, Prof. X. Roy
Department of Chemistry, Columbia University, New York, NY 10027, USA
E-mail: xr2114@columbia.edu

Prof. Xiaodong Xu
Department of Physics, University of Washington, Seattle, WA 98195, USA

Dr. Y.-S. Chen
NSF's ChemMatCARS, University of Chicago, Chicago, Il 60439, USA

$\Delta$ These authors contributed equally

### Single crystal CrSBr synthesis

CrSBr single crystals were synthesized using a modified chemical vapor transport approach adapted from the original report by Beck[1]. Disulfur dibromide and chromium metal were added together in a 7:13 molar ratio to a fused silica tube approximately 35 cm in length, which was sealed under vacuum and placed in a three-zone tube furnace. The tube was heated in a temperature gradient (1223 to 1123 K) for 120 hours. CrSBr grows as black, shiny flat needles along with $CrBr_3$ and $Cr_2S_3$ as side products. CrSBr crystals were cleaned first by washing in warm pyridine, water, and acetone, and then by mechanical exfoliation to ensure no impurities remained on the surface.

### Transport Device fabrication

Exfoliated CrSBr crystals were bonded to a 16-pin DIP socket using low temperature non-conducting epoxy (Loctite EA 1C). Direct electrical connections to the sample were made by hand with silver paint (Dupont 4929N) and 25 μm diameter 99.99% gold wire.

### Transport measurements

Longitudinal resistance was measured in a two-terminal configuration using an SRS830 lock-in amplifier to source voltage and measure current using a 17.777 Hz reference frequency. Hall measurements were performed in a four-terminal configuration using SRS830 lock-in amplifiers to source voltage, measure current, and measure the Hall voltage using a 17.777 Hz reference frequency. Variable temperatures between 1.6 K and 300 K and magnetic fields between -9 T and 9 T were achieved in a Janis pumped-$^4$He cryostat. Sample temperature equilibrium was checked by monitoring sample resistivity for stability over time at a fixed temperature. Hysteresis in the superconducting magnet due to trapped fields was measured by identifying the zero-field shift from the forward and backward field scans measured in the non-magnetic state (at $T$ = 300 K). This hysteresis was accounted for in all presented magnetoresistance measurements. All transport measurements were repeated for multiple samples.

**SQUID measurements**

DC magnetic susceptibility was measured in a Cryogenic R-700 X SQUID magnetometer. Sample masses were on the order of 1 mg and were prepared in ambient conditions inside of gel capsules which were subsequently punctured to ensure no air would remain in the sample upon evacuation. Temperature and magnetic field were carefully monitored for stability at each data point.

**Scanning Tunneling Microscopy and Spectroscopy**

Scanning tunneling microscopy (STM) was performed on freshly cleaved bulk CrSBr crystals using our home-made variable temperature STM at $T = 150$ K in an ultrahigh vacuum chamber (base pressure $< 4.0 \times 10^{-10}$ torr). STM topographical images were obtained in constant current mode ($V_{bias} = 2$V, $I_{tunneling} = 2$ pA) with electrochemically etched tungsten tips. To avoid tip artifacts, each STM tip was calibrated on a clean Au (111) surface before all measurements. All tips were verified to be atomically sharp. We obtained the differential conductance (dI/dV) spectra with a lock-in amplifier while keeping the tip fixed above the surface with the feedback loop off.

**Photoluminescence spectroscopy**

Photoluminescence spectroscopy was performed with a 633 nm HeNe laser with Princeton Instruments SpectraPro HRS-300 spectrograph and PyLoN-IR camera. Samples were mounted in a gas-tight cell with a UVFS window and a nitrogen atmosphere. CrSBr crystals were cleaned by cleaving their surfaces using Scotch tape before the measurements. Measurements were performed in a microscopic setup and an on-sample power was maintained at 200 μW.

**Chemical and structural analysis**

*Scanning electron microscopy and energy dispersive x-ray spectroscopy*

Scanning electron micrographs were collected on a Zeiss Sigma VP scanning electron microscope (SEM). Energy dispersive x-ray spectroscopy (EDX) of the CrSBr crystals was performed with a Bruker XFlash 6 | 30 attachment. Spectra were collected with a beam energy of 15 kV. Elemental compositions and atomic percentages were estimated by integrating under the characteristic spectrum peaks for each element using Bruker ESPRIT 2 software.

*Single crystal x-ray diffraction - Columbia*

Single crystal x-ray diffraction data were collected on an Agilent SuperNova diffractometer using mirror-monochromated Mo K$\alpha$ radiation. Each crystal was mounted under oil at room temperature using a MiTeGen MicroMount. Data collection and unit cell determination was performed in CrysAlisPro. Samples were matched against previous crystallographic characterizations.[1,2]

*Synchrotron low temperature single crystal x-ray diffraction – NSF'sChemMatCARS*

A single crystal with dimension of 10x20x12 $\mu m^3$ was mounted on the tip of a glass fiber. Data were collected at x-ray energy 30 keV (0.41328Å) using Huber 3 circles diffractometer with kappa angle offset 60° and equipped with Pilatus3X 1M (CdTe) detector at temperature of 50 and 15 K. The distance between the detector and crystal was 130 mm. A total of 1440 frames were collected at two θ-angles sitting at 0° follow up two different ω-angles: -180 °; kappa: 0 ° and ω-angles:-200˚; kappa: 30˚, respectively. The data were collected while the φ-angle was scanned over the range of 360˚ using shutterless mode. A user-friendly data collection software was used. Pilatus standard CBF frames were converted to Bruker sfrm format. Data integrations were performed with APEX II suite software. The reduction of data was conducted with SAINT

v.8.32B and SADABS v.2013 programs included in the APEX suite. The structure solution and refinement were carried out with SHELX software using the XPREP utility for the space group determination, and the XT and XL programs for the structure solution and refinement, respectively.

*Raman spectroscopy*

Raman spectra were acquired in a Renishaw inVia micro-Raman microscope using a 532 nm wavelength laser. A 50x objective was used with a laser spot size of ~2 - 3 µm and a laser power of 200 µW. Spectra were acquired for 30 seconds with a grating of 2400 gr mm$^{-1}$. 10 spectra were acquired and averaged after removing noise from cosmic background radiation and the detector.

**Figures:**

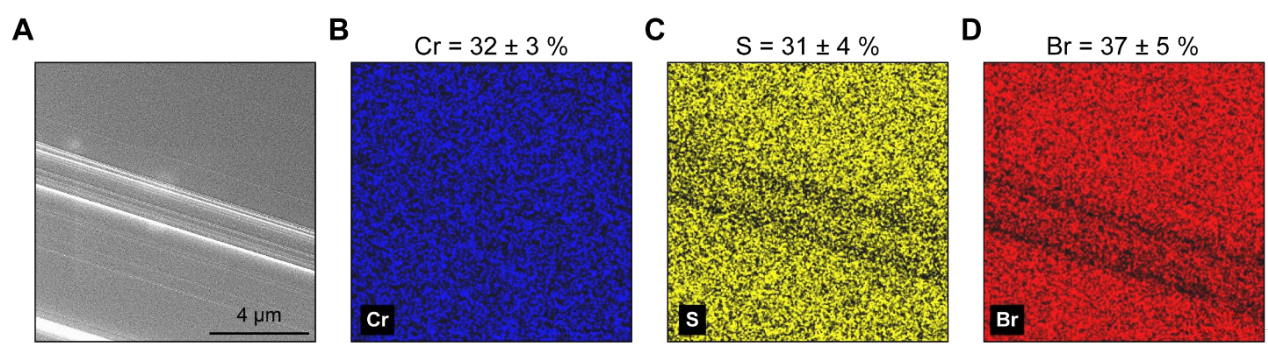

**Figure S1. Chemical composition of CrSBr.** (**A**) SEM image of bulk CrSBr as grown after cleaning. (**B-D**) Corresponding EDX chemical maps for Cr (**B**), S (**C**), and Br (**D**). Extracted chemical percentages are given above the EDX maps.

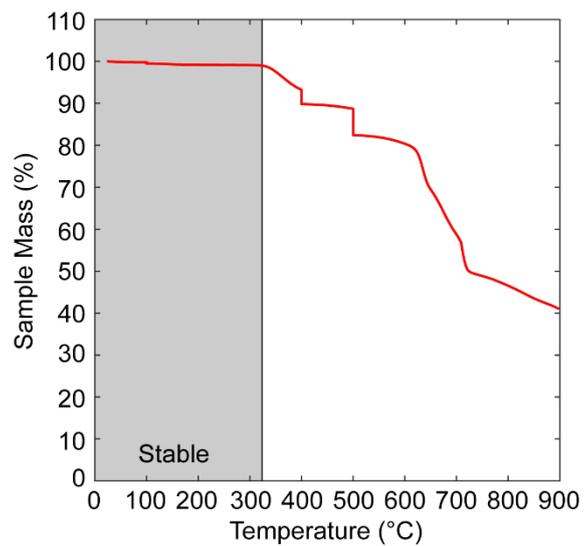

**Figure S2. Thermogravimetric analysis of CrSBr.** Percentage mass of CrSBr crystals versus temperature in a $N_2$ atmosphere. Grey region denotes the temperature range over which the crystals are stable.

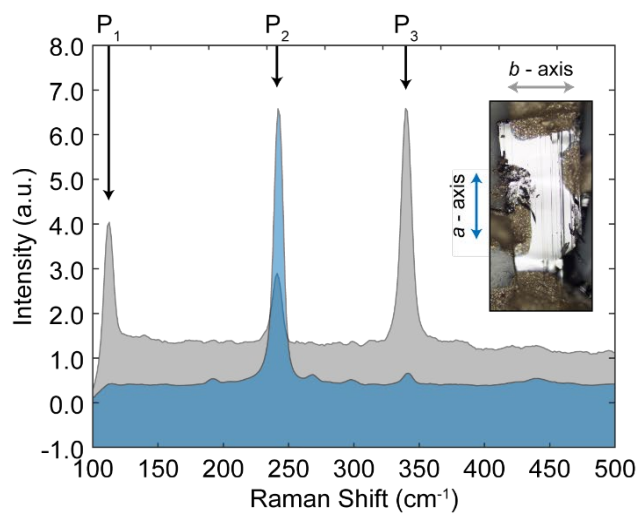

**Figure S3. Assignment of crystal axes through optical imaging and Raman spectroscopy.** Raman signal intensity versus wave number for light polarized parallel to the *b*-axis (grey curve) and *a*-axis (blue curve). Inset: optical image of a bulk transport device with the light polarization marked relative to the physical crystal axes. Three distinct Raman modes ($P_1$, $P_2$, and $P_3$) are denoted. $P_1$ and $P_3$ are only observed with a laser polarization aligned parallel to the *b*-axis, which we use to confirm the orientation of the crystals before mounting into the cryostat.

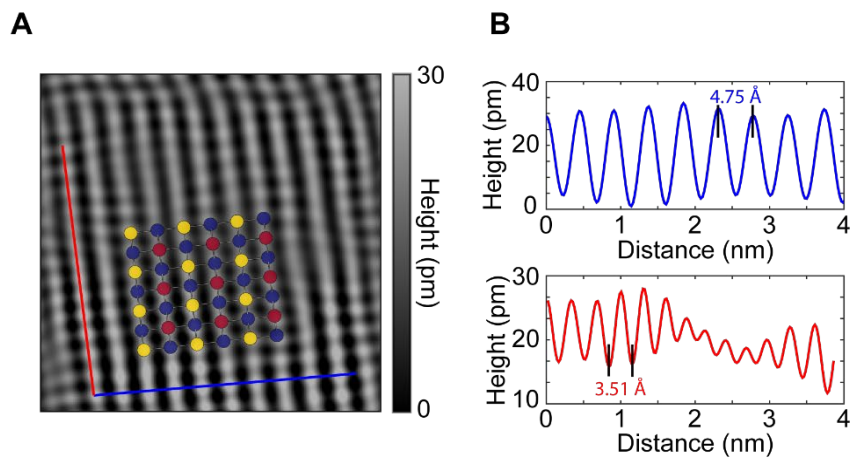

**Figure S4. Atomic structure of CrSBr.** (**A**) STM topography of bulk CrSBr as viewed along the *c*-axis. The crystal structure is overlaid to emphasize the lattice structure. (**B**) Line cuts of (**A**) along the blue (**B:top**) and red (**B:bottom**) lines shown in (**A**). The extracted lattice parameters are given in the insets of **B:top** and **B:bottom**.

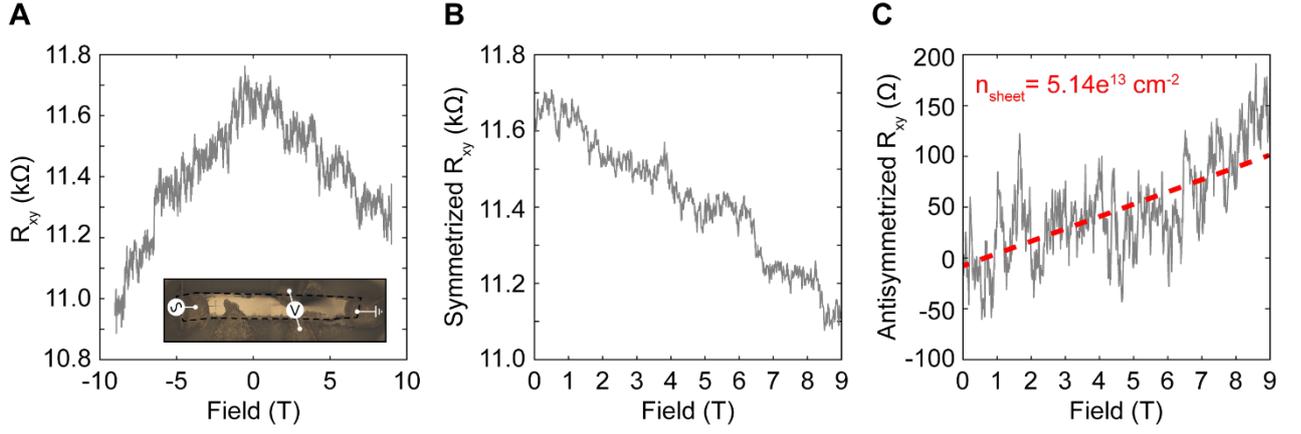

**Figure S5. CrSBr electronic carrier density at 300 K.** (**A**) Raw Hall ($R_{xy} = \frac{V_{xy}}{I}$) resistance versus magnetic field. Inset shows an image of a transport device with the measurement schematic overlaid. (**B**) Symmetrized Hall resistance versus magnetic field $R_{xy-sym} = \frac{R_{xy}(+B)+R_{xy}(-B)}{2}$. (**C**) Anti-symmetrized Hall resistance versus magnetic field $R_{xy-asym} = \frac{R_{xy}(+B)-R_{xy}(-B)}{2}$. A linear fit is shown as a dashed red line. The extracted carrier density per sheet is given in the inset.

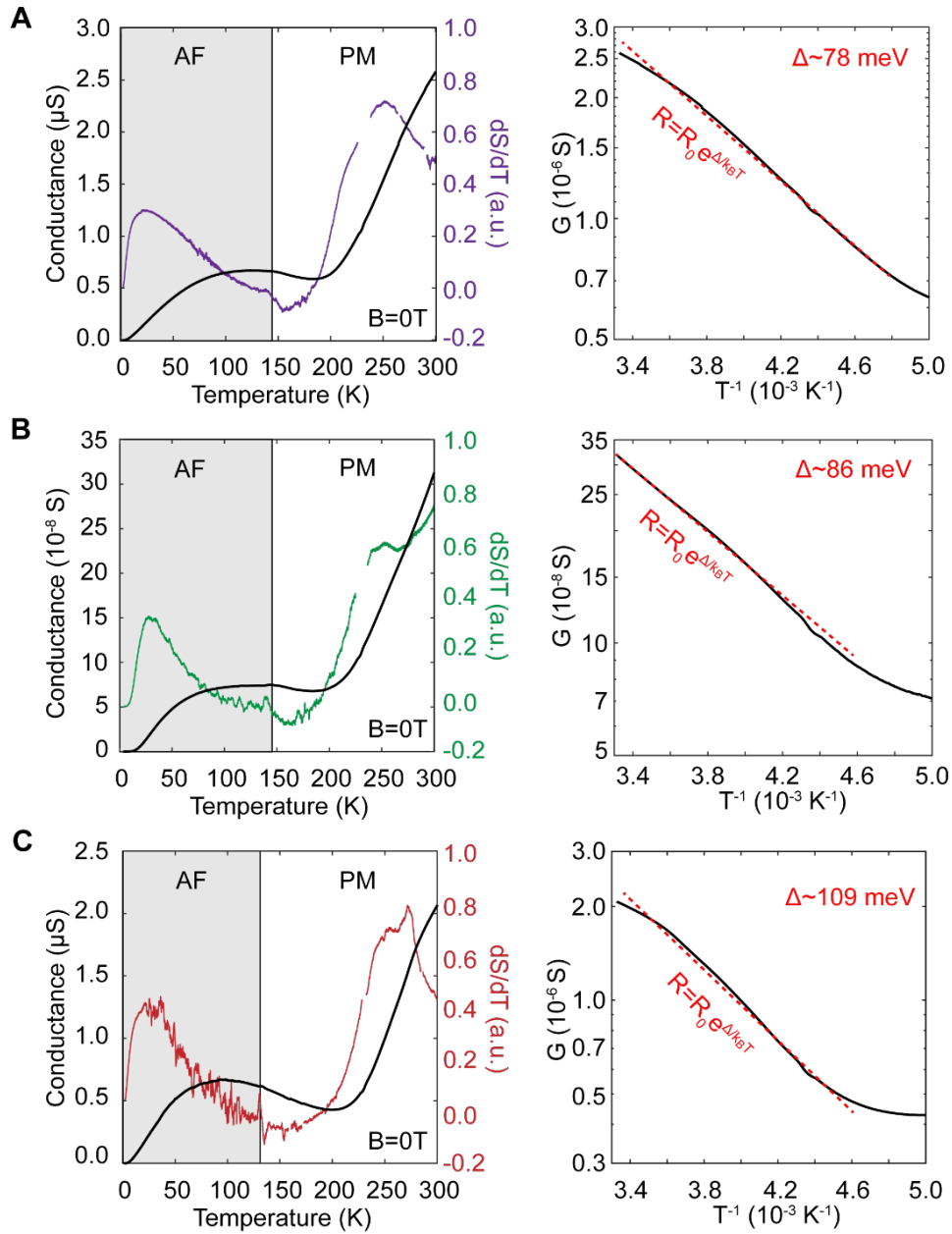

**Figure S6. Repeatability of zero-field transport behavior.** (**A-C: Left**) Conductance (left axis) and the derivative of conductance (right axis) versus temperature for 3 additional devices. The AF and PM phases are denoted with grey and white regions, respectively. Extracted Néel temperatures are 147±7 K, 148±8 K, 133±5 K for (**A**), (**B**), and (**C**) respectively. (**A-C: Right**) Conductance on a log scale versus inverse temperature. Linear fits to a thermal activation model are given by red dashed lines. The extracted transport gaps are given in the inset.

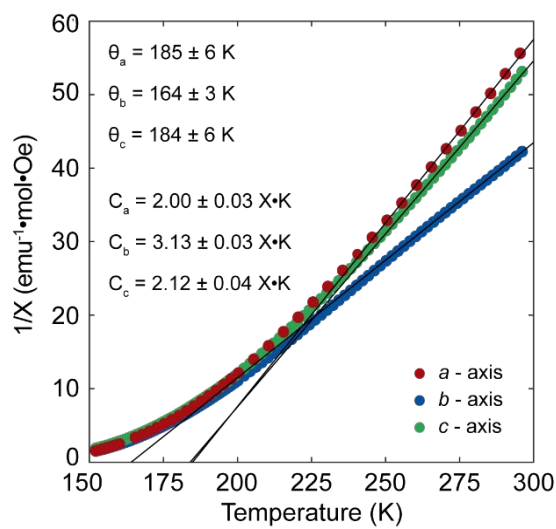

**Figure S7. Curie-Weiss fitting of the magnetic susceptibility.** Inverse magnetic susceptibility versus temperature along the *a*-axis (red dots), *b*-axis (blue dots), and *c*-axis (green dots). Linear fits to the paramagnetic regime are given by solid black lines. Extracted Curie and Weiss constants are given in the inset.

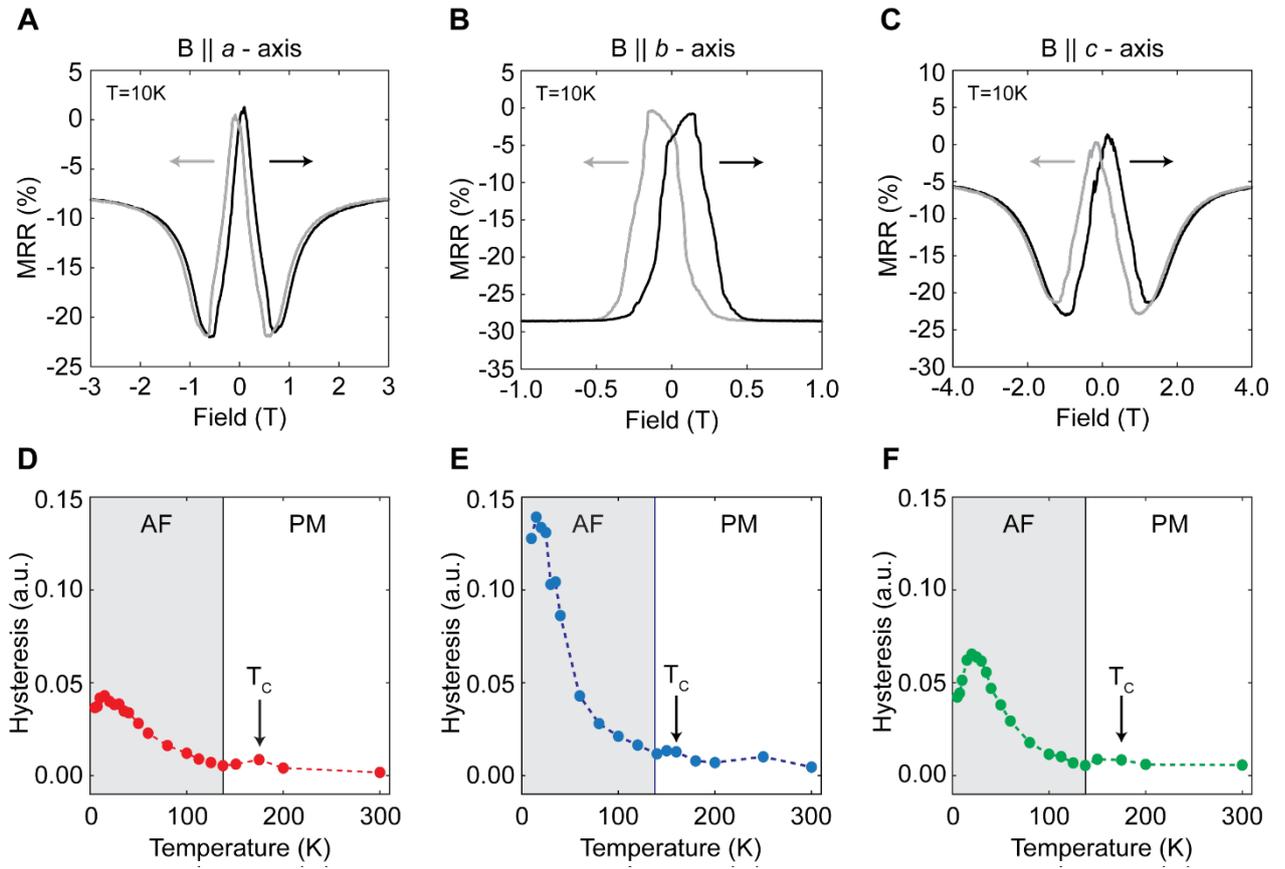

**Figure S8: Hysteresis and magnetic ordering above the Néel temperature.** (**A-C**) Magnetoresistance ratio versus magnetic field at 10 K with the field oriented along the *a*-axis (**A**), *b*-axis (**B**), and *c*-axis (**C**). Both forward (solid black line) and backward (solid grey line) magnetic field scans are presented. (**D-F**) Hysteresis in magnetoresistance, defined as the integrated absolute difference between the magnetoresistance ratio measured while sweeping the magnetic field forward minus the magnetoresistance ratio measured while sweeping the magnetic field backward ($H = \int |MRR(B_\rightarrow) - MRR(B_\leftarrow)| dB$), versus temperature with the magnetic field oriented along the *a*-axis (**D**), *b*-axis (**E**), and the *c*-axis (**F**). Magnetic phases as identified by the SQUID measurements are overlaid.

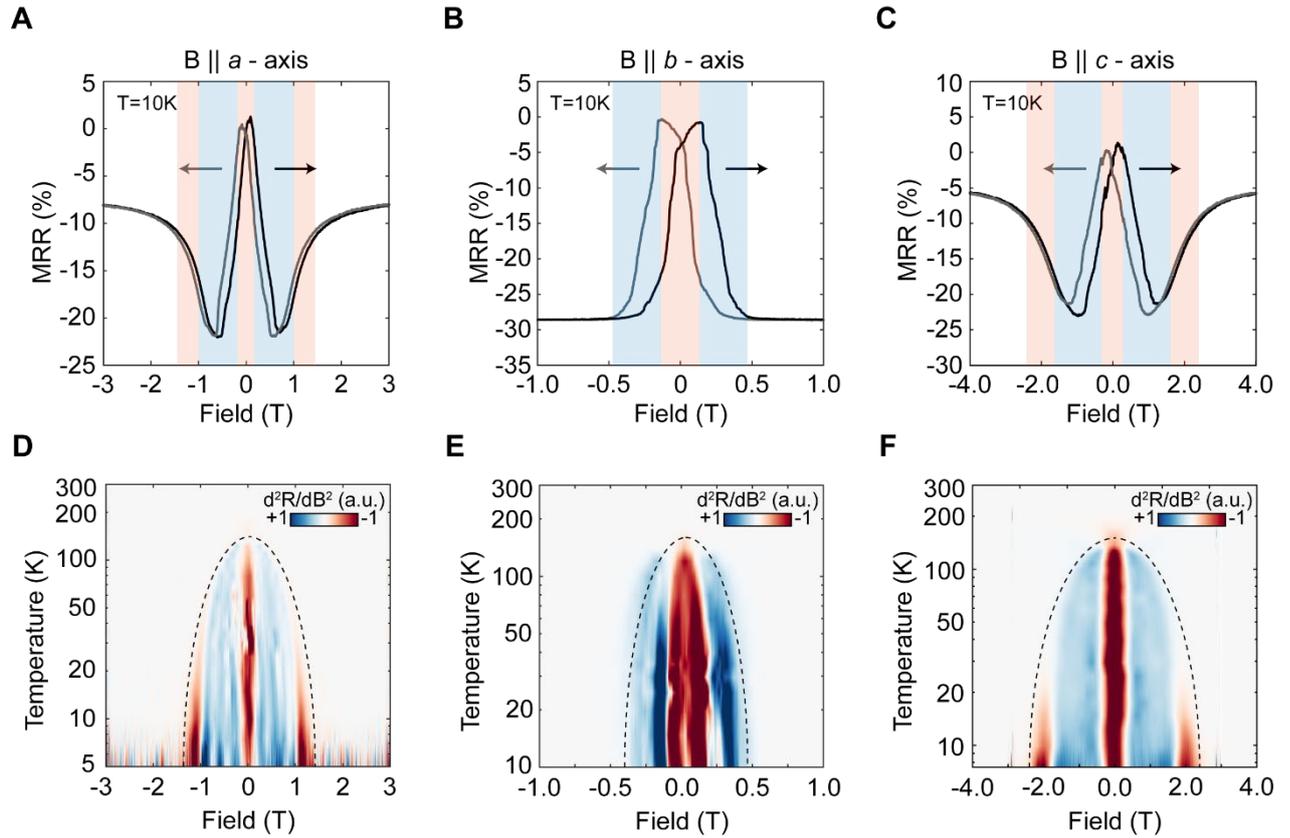

**Figure S9: Extracting saturation magnetic fields versus temperature.** (**A-C**) Magnetoresistance ratio versus magnetic field at 10 K with the field oriented along the *a*-axis (**A**), *b*-axis (**B**), and *c*-axis (**C**). Both forward (solid black line) and backward (solid grey line) magnetic field scans are presented. (**D-F**) Second derivative of the magnetoresistance ratio versus magnetic field and temperature with the field oriented along the *a*-axis (**D**), *b*-axis (**E**), and *c*-axis (**F**). The corresponding blue and red features are labeled in **A-C**. Only the forward field scans are presented in **D-F**. In **D-F**, a dashed black line is plotted as a guide to the eye tracking the saturation field versus temperature.

**Table S1. Selected crystallographic data.**

| T (K) | 15 | 50 | 100 | 294 |
|---|---|---|---|---|
| Formula | CrSBr | CrSBr | CrSBr | CrSBr |
| MW | 163.87 | 163.87 | 163.97 | 163.97 |
| Space Group | Pmmn | Pmmn | Pmmn | Pmmn |
| a (Å) | 3.5126 | 3.5138 | 3.5043 | 3.5029 |
| b (Å) | 4.7457 | 4.7484 | 4.7379 | 4.7631 |
| c (Å) | 7.9171 | 7.9271 | 7.9069 | 7.963 |
| α (°) | 90 | 90 | 90 | 90 |
| β (°) | 90 | 90 | 90 | 90 |
| γ (°) | 90 | 90 | 90 | 90 |
| V (Å$^3$) | 131.98 | 132.26 | 131.28 | 132.86 |
| Z | 2 | 2 | 2 | 2 |
| $\rho_{calc}$ (g cm$^{-3}$) | 4.126 | 4.117 | 4.148 | 4.099 |
| l (Å) | 0.3936 | 0.3936 | 0.71073 | 0.71073 |
| $2q_{min}, 2q_{max}$ | 5.542, 51.848 | 2.846, 60.142 | 10.032, 58.958 | 9.974, 52.396 |
| Nref | 4347 | 744 | 1685 | 223 |
| R(int), R(s) | 0.0312, 0.0233 | 0.0446, 0.0341 | 0.0735, 0.0406 | 0.0624, 0.0837 |
| μ (mm$^{-1}$) | 3.762 | 3.754 | 19.976 | 19.738 |
| Data | 661 | 744 | 219 | 142 |
| Restraints | 0 | 0 | 0 | 0 |
| Parameters | 14 | 14 | 13 | 13 |
| $R_1$ (obs) | 0.0225 | 0.0411 | 0.0674 | 0.0779 |
| $wR_2$ (all) | 0.0787 | 0.1741 | 0.1959 | 0.1861 |
| S | 1.334 | 1.249 | 1.486 | 1.101 |

**Table S2.** Curie-Weiss fit parameters

| Axis | Curie constant (χ•K) | Θ (K) |
|---|---|---|
| *a* | 2.00 ± 0.03 | 185 ± 6 |
| *b* | 3.13 ± 0.03 | 164 ± 3 |
| *c* | 2.12 ± 0.04 | 184 ± 6 |

**Supporting Information References:**